# Self-solidifying active droplets showing memory-induced chirality

*Kai Feng,[#] José Carlos Ureña Marcos,[#] Aritra K. Mukhopadhyay, Ran Niu,[*] Qiang Zhao, Jinping Qu, and Benno Liebchen*

K. Feng, R. Niu, Q. Zhao, J. P. Qu

Key Laboratory of Material Chemistry for Energy Conversion and Storage, Ministry of Education, School of Chemistry and Chemical Engineering, Huazhong University of Science and Technology, Wuhan 430074, China

Email: niuran@hust.edu.cn

J. C. Ureña Marcos, A. K. Mukhopadhyay, B. Liebchen

Institut für Physik Kondensierter Materie, Technische Universität Darmstadt, Darmstadt, Germany

**Abstract:** Most synthetic microswimmers do not reach the autonomy of their biological counterparts in terms of energy supply and diversity of motions. Here we report the first all-aqueous droplet swimmer powered by self-generated polyelectrolyte gradients, which shows memory-induced chirality while self-solidifying. An aqueous solution of surface tension–lowering polyelectrolytes self-solidifies on the surface of acidic water, during which polyelectrolytes are gradually emitted into the surrounding water and induce linear self-propulsion via spontaneous symmetry breaking. The low diffusion coefficient of the polyelectrolytes leads to long-lived chemical trails which cause memory effects that drive a transition from linear to chiral motion without requiring any imposed symmetry breaking. The droplet swimmer is capable of highly efficient removal (up to 85%) of uranium from aqueous solutions within 90 min, benefiting from self-propulsion and flow-induced mixing. Our results provide a route to fueling self-propelled agents which can autonomously perform chiral motion and collect toxins.

**Keywords**: droplet swimmer, memory-induced chirality, self-solidifying, Marangoni flow, polyelectrolyte

## 1. Introduction

Synthetic active matter comprises agents which can propel themselves in an unbiased environment, such as Janus colloids,[1-7] motile droplets[8-15] or microflyers.[16] These (micro)swimmers can act as motors which operate without any moving parts or electronic ingredients and can carry drugs and other cargo even at subcellular scales.[17-19] There has been vigorous research regarding the design of synthetic swimmers in the past decade. In most cases, these swimmers essentially contain a compartment filled with low surface tension, water-immiscible organic fluids which serve as the fuel to drive their motion.[20-22] Alternatively, swimmer motion can be driven by chemical reactions which produce gaseous products or ionic gradients, such as the oxidation of hydrogen peroxide.[23,24] Nevertheless, neither the use of organic solvents nor toxic chemical reactions are relevant for practical environmental remediation applications. In this regard, there is an urgent need for a paradigm shift in the way of incorporating the driving force of synthetic swimmers which allows for the design of *all-aqueous* microswimmers suitable for mass transfer and exchange with the surrounding environment, as required e.g., for water remediation.

Another key aspect concerning prospective applications of synthetic swimmers is the diversity of motions that the employed agents can perform. In this respect, synthetic swimmers still lag behind biological swimmers. According to the type of motion, swimmers can be broadly classified as linear or chiral swimmers. While linear swimmers feature a characteristic "self-propulsion velocity" and move linearly or ballistically (in competition with fluctuations), chiral swimmers show a characteristic frequency leading to chiral trajectories with a circular or helical shape.[25-30] Most swimmers belong to the former class, but chiral swimmers offer advantages for problems like optimal surface coverage,[31] which are important for the survival of living microswimmers and possible applications of synthetic swimmers, such as the cleaning of wastewater. Importantly, chirality also plays a crucial role in the collective behavior of the swimmers,[32-34] and can lead to patterns such as arrays of rotating microflocks,[35,36] vortex arrays[37-39] and synchronized colloidal cogwheels.[40] This rich phenomenology contrasts with a relatively limited list of mechanisms available to create chiral self-propulsion. The most obvious mechanism depends on chiral particle shapes, which couple rotational and translational degrees of freedom of a swimmer, as illustrated by L-shaped colloids[27,41] or droplets either involving chiral molecular motors,[42] helical director fields[43] or undergoing phase separation.[44] Alternatively, chiral self-propulsion can emerge in viscoelastic fluids through a delay in the response of the medium to the motion of the swimmer.[45] Finally, a third possible route to chirality uses hydrodynamic interactions with walls or interfaces.[12,26,46-49] Overall, all the known routes to chirality require asymmetric particle shapes/director fields[27,41,43] or nontrivial environments comprising viscoelastic fluids[45], interfaces or external time-dependent driving forces.[50,51] (Note that chirality can also emerge at a many-body level from anisotropic interactions.[40,52])

In this work, we demonstrate a new type of all-aqueous isotropic polyelectrolyte droplet swimmer, which creates (chiral) self-propulsion by self-generated, long-lived chemical gradients in a simple and isotropic environment, similar to that of oil-in-water droplets.[8,20] The droplet swimmer, which consists of two polyelectrolytes that undergo pH-dependent complexation, gradually self-solidifies when it is in contact with acidic water. This self-solidifying enables the isotropic release of a surface tension–lowering polyelectrolyte, which leads to self-propulsion through spontaneous symmetry breaking. That is, although the swimmer shape and the release of PSS are isotropic, any small perturbation of the isotropic configuration creates an asymmetry in the surface tension and flow field profiles, which is reinforced in suitable parameter regimes (positive feedback loop),[8,53] leading to self-propulsion. The latter is observable in our experiments over timescales of at least 20 min. The released polyelectrolytes diffuse slowly and form long-lived trails which influence the motion of the swimmer. These self-interactions of the swimmer with its own trajectory, or history, which represent memory effects, induce a continuous reorientation and thus chiral motion of the swimmer. Furthermore, the droplet swimmer acts as a "cleaning robot" in uranium-polluted water, reaching highly efficient removal of uranium (85%) thanks to self-propulsion and flow-enhanced mixing. Our results exemplify an alternative route to self-powered swimmers, such as droplets and hydrogels, that experience pronounced self-induced memory effects, as well as a method of managing polluted water. These memory effects constitute an alternative mechanism for chiral swimmer motion, which could inspire the design of microswimmers capable of optimal surface coverage,[31] negative viscotaxis[54,55] or swimming in regions where wastewater treatment is needed.

## 2. Results
### 2.1. Self-solidifying droplet swimmers

We prepared a homogeneous aqueous solution containing poly(ethyleneimine) and poly(sodium 4-styrenesulfonate), to which we refer here as the PEI/PSS solution.[56] Noteworthily, the pH of the PEI/PSS solution is 13, at which both protonation of amine groups ($-NH_2$ in PEI) and PEI/PSS complexation are very weak. Therefore, the PEI/PSS solution is homogeneous without any precipitation. When a droplet of the PEI/PSS solution (20 μL) is placed on an acidic water solution (pH 1.25, Figure 1a), surprisingly, instead of dissolving in the underlying water, it spontaneously self-propels from the center of the Petri dish towards the rim (Figure 1b and Supplementary Movie S1). Once the droplet has reached the dish rim, it turns and moves ballistically along the confining walls, while keeping a characteristic distance to the boundary. Remarkably, after a few hundred seconds, the droplet suddenly leaves the boundary of the Petri dish and moves along a chiral trajectory (in the sense that it can be distinguished from its mirror image), the characteristic radius of which is much smaller than that of the Petri dish (Figure 1b and Supplementary Movie S2). After yet another few hundred seconds, the character of the droplet motion changes again and leads to a persistent localized motion around the center of the Petri dish (Figure 1b and Supplementary Movie S3). While self-propelling, the droplet reaches a peak velocity of about 6.3 cm s$^{-1}$ before gradually slowing down (Figure 1c) and covers a total distance of about 14 m within 1000 s (inset of Figure 1c). The concentration of polyelectrolytes influences the velocity and total distance covered by the droplet. In particular, we find the best results for 25 wt% solids content (Figure S1).

During the water-surfing movement, the PEI/PSS swimmer gradually changes from transparent to opaque, forming a UFO-like shape (Figure 1d). That is, the periphery of the droplet, i.e., the air-water-droplet interface, turns opaque first, and this opaqueness gradually advances towards the droplet center. Since the surrounding water solution is acidic (pH 1.25), protons diffuse into the droplet across the water-droplet interface and trigger PEI/PSS complexation (Figure 1e, solid arrows). Due to the rapid kinetics of polyelectrolyte complexation,[56,57] the droplet quickly solidifies at its boundary, which prevents the droplet from dissolving in the underlying water. Although the swimmer is different from classic droplets due to the formation of a thin solid boundary, we refer to it as a droplet as there is still mass exchange between the droplet and its surroundings.

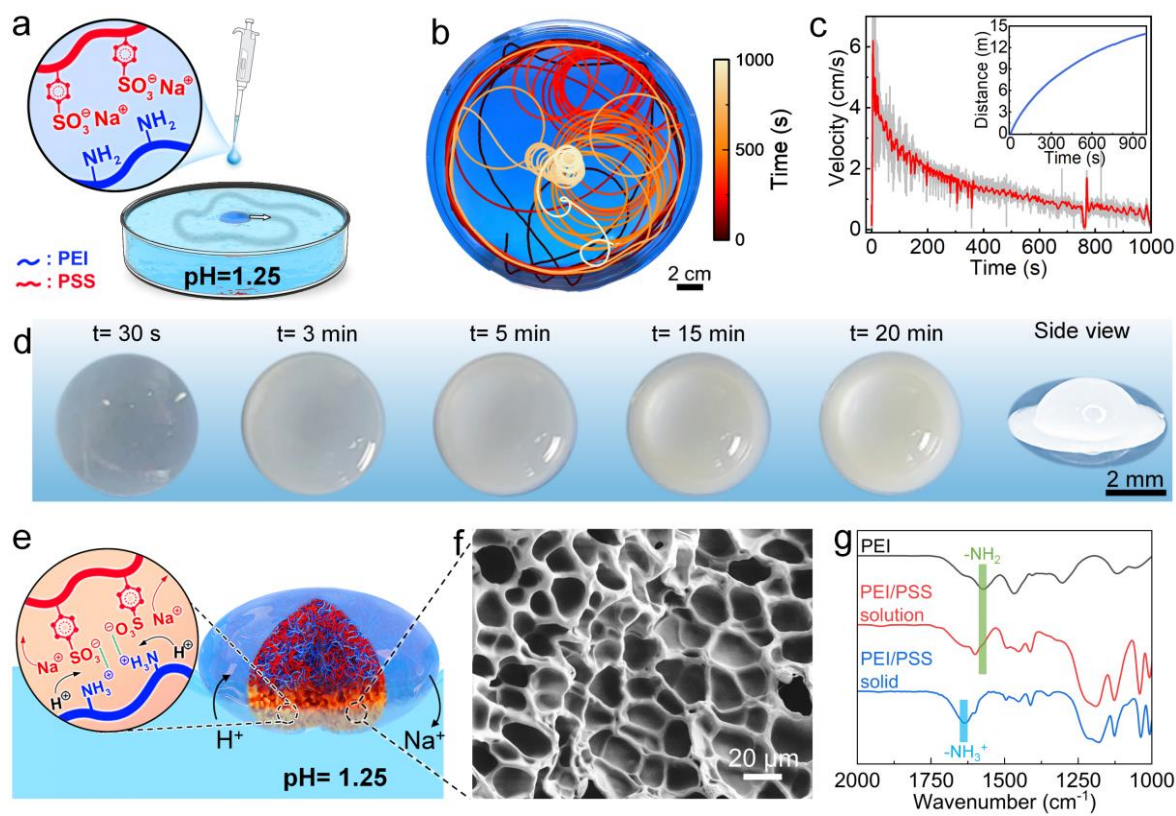

**Figure 1. Design and characterization of self-solidifying droplet swimmers.** (a) Schematic of the droplet motion and photograph of the PEI/PSS solution (25 wt% solids content). (b) Typical droplet trajectory in a Petri dish of radius $R = 10$ cm. (c) Droplet speed corresponding to the trajectory shown in (b). Inset: total distance traveled by the droplet. (d)

Onset and progress of droplet self-solidifying. (e) Schematic of the self-solidifying of PEI/PSS at the air-water-droplet interface. (f) Microstructure of the solidified region of the droplet. (g) Fourier-transform infrared spectroscopy (FTIR) spectra of PEI, the PEI/PSS solution and solidified PEI/PSS.

The solidified region of the droplet is porous (Figure 1f), which is consistent with the phase inversion caused by polyelectrolyte complexation.[58] These large pores facilitate the diffusion of PEI and PSS from the unsolidified region into water, which was verified by the characteristic UV absorption of PSS (262 nm) detected in the surrounding water (Figure S2). Furthermore, the FTIR spectra show a shift from 1570 cm$^{-1}$ to 1635 cm$^{-1}$, confirming the protonation of amine groups triggered by the acidic water solution (Figure 1g). After placing a PEI/PSS droplet on water (pH 1.25) for 1.5 h, the total droplet mass loss is 22.1 wt% (Figure 2a), with the release rate of PSS decreasing with time and eventually leveling off (Figure 2b). The molar ratio of PSS to PEI influences the mass loss of PSS, which reaches a minimum at 1:1 (Figure S3a), as well as the velocity and self-propulsion time of the droplet (Figure S3b). Moreover, both the average speed and the total distance traveled by the droplet decrease with the concentration of ethanol in the underlying water (Figure 2c). This is probably because the surface tension of the underlying water is reduced due to the addition of ethanol (Figure 2d), whose surface tension (22.3 mN m$^{-1}$) is lower than that of the acidic water solution (72.8 mN m$^{-1}$).

We explored the influence of the confining boundary on the droplet motion by performing experiments in square (50 × 50 cm), hexagonal (side 20 cm) and rectangular (30 × 40 cm) vessels. Similar to the results in the circular Petri dish (R=10 cm), the droplet moves towards the rim and then describes a chiral trajectory along the path it followed to first reach the boundary (Figure 2e, S4a and S4b). Chiral trajectories were also obtained in a tank (100 × 60 cm, Figure S4c). These observations indicate that chiral self-propulsion of the droplet swimmer does not require a certain kind of symmetry in the confining boundary. However, in strong one-dimensional confinement (~ 7 droplet diameters), chiral motion does not occur; rather, the droplet moves along the channel performing directed motion (Figure S4d).

Remarkably, the release of PSS can be exploited to create surface tension gradients for self-propulsion of other objects, such as heart-shaped poly(acrylic acid- acrylamide) or gelatin hydrogels (Figure 2i and S5). However, the diffusion of PSS from hydrogels is relatively inhibited due to their intrinsic nanopores, resulting in a shorter self-propulsion time and a lower velocity than those of the PEI/PSS droplet swimmer. Importantly, when PSS of lower molecular weight (70,000), and therefore higher diffusion coefficient (48.5 μm$^2$ s$^{-1}$), is used in the droplet, chiral motion does not occur (Figure S6), which is in line with our analysis of the mechanism underlying chiral motion, which requires a localized (nonuniform) distribution of PSS, as we discuss further below.

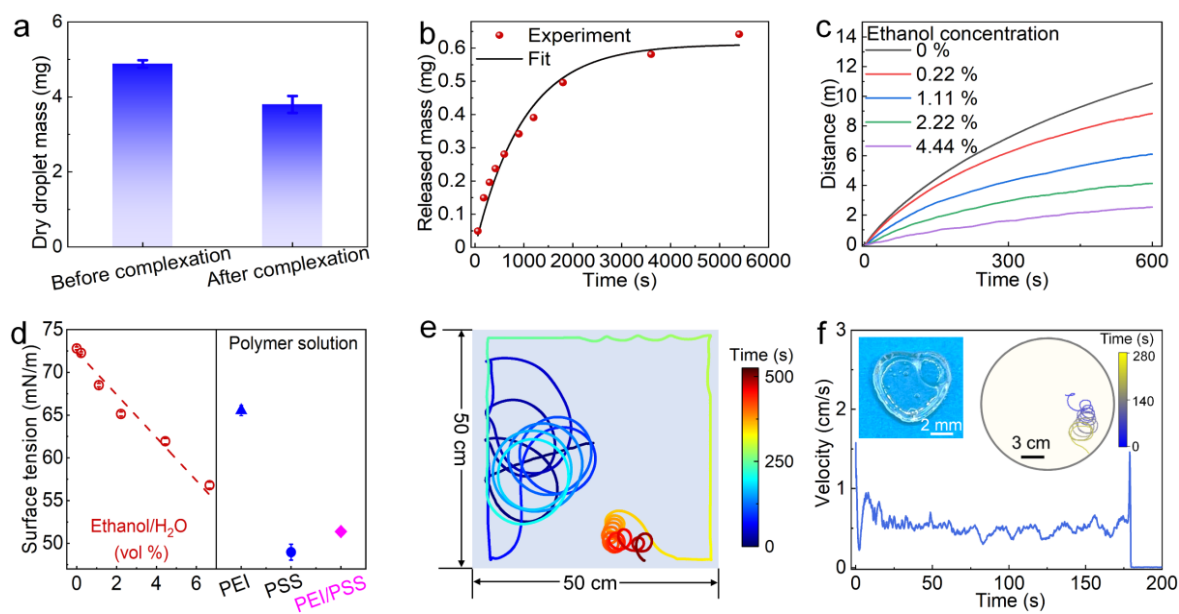

**Figure 2. Characteristics of droplet motion.** (a) Dry droplet mass before and after complexation in pH 1.25 water. (b) Total released mass of PSS versus time. The black line corresponds to a fit of the form $\int_0^t ae^{-bt'}dt' = \frac{a}{b}(1-e^{-bt})$, which yields $a = 0.037$ mg min$^{-1}$ and $b = 0.060$ min$^{-1}$. (c) Total distance traveled by the droplet swimmer in ethanol/water solutions of different volume fractions. (d) Surface tension of various solutions. (e) Droplet trajectory in a square glass vessel of side 50 cm. (f) Velocity of a poly(acrylic acid-acrylamide) hydrogel with embedded PSS molecules. Insets: photograph of the heart-shaped hydrogel (left) and its trajectory (right).

## 2.2. Self-solidifying induces Marangoni flows

Let us now address the driving force for self-propulsion of the droplet and why its motion suddenly changes from ballistic to chiral. As described above, the self-solidifying process of the droplet is accompanied by a continuous emission of PSS into the surrounding water (Figure 2a and S2). Importantly, the emitted PSS lowers the surface tension in the vicinity of the droplet (Figure S7), which in turn induces long-ranged Marangoni flows at the air-water interface (Figure 3a,b). The resulting overall flow profiles are three-dimensional (Figure 3a and S7), as expected due to the incompressibility of the acidic water solution. Remarkably, when fixing the droplet with a needle at the air-water interface so that it acts as a pump (Figure S7), the flow at the air-water interface decays in an essentially linear fashion with increasing radial distance. This is consistent with our simulations based on a 2D model (Figure 3c; see Model in the *Methods* section).

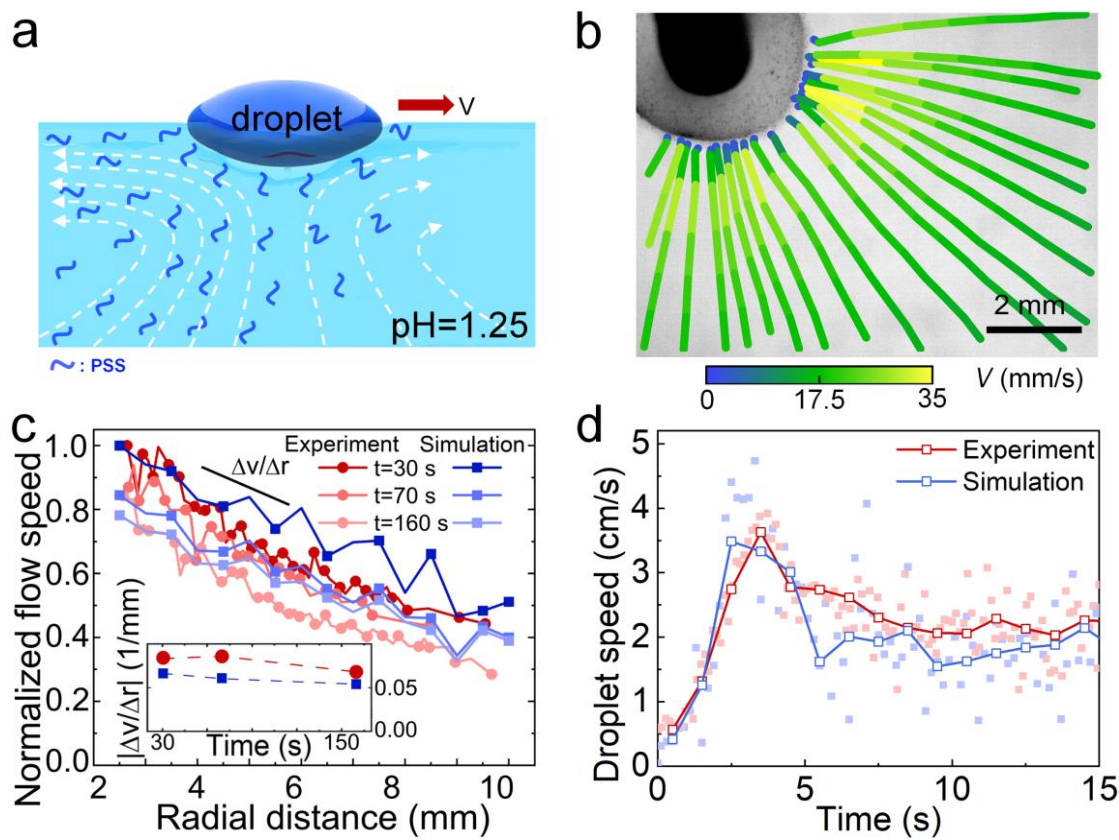

**Figure 3. Marangoni flow and droplet speed.** (a) Schematic for the droplet swimmer showing the asymmetric distribution of the released PSS (spontaneous symmetry breaking,

see main text). The asymmetry in the PSS distribution evokes a surface tension gradient, which in turn leads to an asymmetric flow (Marangoni effect). The asymmetric flow results in directed motion of the droplet along the air-water interface. (b) Top view of the fluid streamlines around the droplet when it acts as a fixed pump. (c) Radially averaged flow speed at the air-water interface. Inset: change of the slope with time. See Figure S7 for a schematic of the setup used in (a-c). (d) Droplet speed far away from any boundaries (details in *Methods*). Solid squares: raw data. Open squares: binned data.

**2.3. Spontaneous symmetry breaking induces self-propulsion**

To understand the onset of ballistic and chiral self-propulsion, we now develop a detailed 2D model describing the dynamics of the droplet (see details in *Methods*). More precisely, we model the self-solidifying droplet as a disk which is suspended on the air-water interface and emits PSS isotropically along the air-water-droplet contact line. The emitted PSS then spreads out diffusively, leading to a characteristic concentration profile. Since they also reduce surface tension, the droplet initially acts as a pump evoking isotropic Marangoni flows along the air-water interface, which in turn advect the polymer.

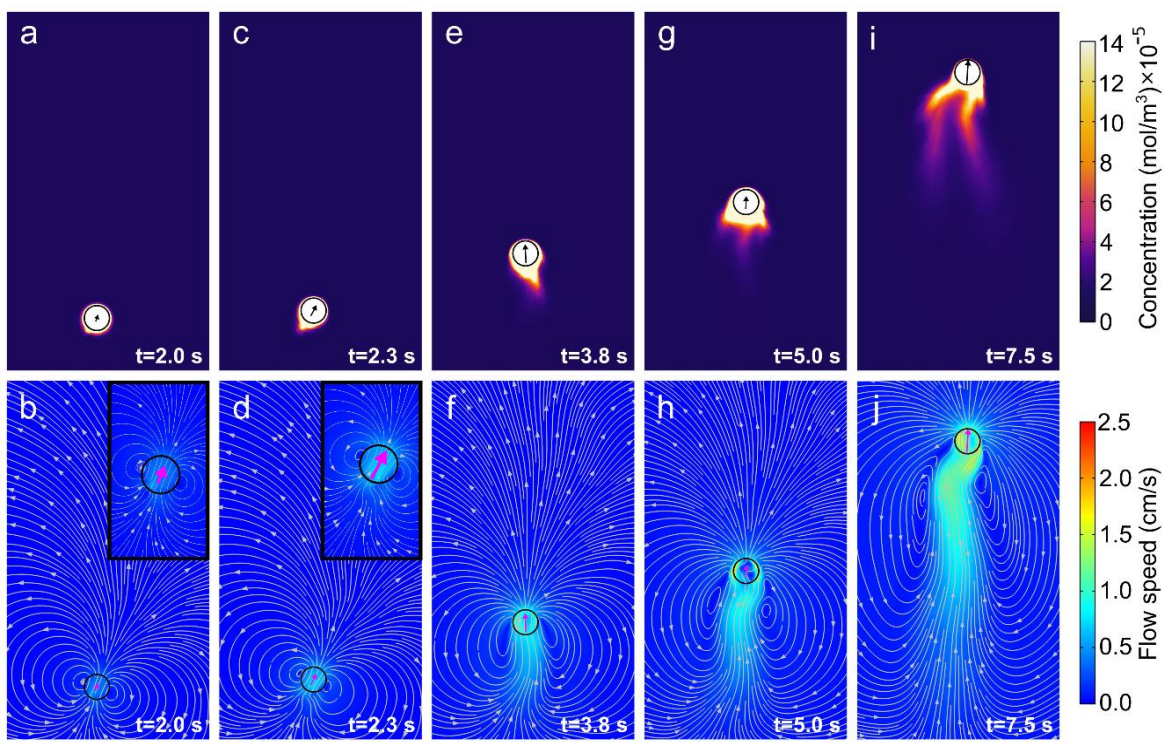

**Figure 4. Self-propulsion mechanism from simulations.** Simulation snapshots of the PSS concentration (a, c, e, g, i) and flow field with streamlines (b, d, f, h, j) close to the droplet swimmer. Arrow lengths are proportional to the droplet speed.

Initially, given the isotropy of the droplet and the resulting Marangoni flows, the droplet does not move. Importantly, however, as time evolves, a positive feedback loop between the PSS distribution and the droplet dynamics sets in which leads to spontaneous symmetry breaking and self-propulsion. The feedback loop works as follows: PSS is isotropically released by the droplet, but any tiny spontaneous fluctuation in the PSS concentration along the droplet boundary (Figure 4a) evokes a slightly asymmetric Marangoni flow (Figure 4b) exerting a weak effective force on the droplet which has direct and indirect contributions (see *Methods* and Ref. [59]). This effective force points towards regions of low PSS concentration (high surface tension). Crucially, since the droplet moves away from regions of high PSS concentration, the droplet motion increases the weak spontaneous asymmetry in the PSS distribution (Figure 4c), which in turn enhances the Marangoni flows (Figure 4d) and the corresponding effective force acting on the droplet. This leads to an enhanced droplet motion which further increases the asymmetry in the PSS concentration (a trail of high PSS concentration develops, Figure 4e,g,i), and hence also the droplet speed (Figure 4f,h,j), up to a certain saturation point. (Note that this feedback loop works as long as the Péclet number is high enough,[7,53] which is ensured in our system by the low diffusivity of PSS (Supplementary Movie S4 and Figure S13).) Our simulations predict the entire buildup of the self-propulsion speed and its slow decay at longer time scales (Figure 3d), which is a consequence of a reduction in the PSS release rate (Figure 2b and S10), in close quantitative agreement with experiments. Remarkably, most of the parameters of our model can be directly extracted from the experiments, except for two fitting parameters whose values are chosen in such a way that the maximum droplet speed matches that of the experiments. The entire shape of the speed curve then follows from this choice, including the position of the maximum (Figure 3d).

**2.4. Transition from ballistic to chiral active motion**

As in experiments, our simulations show that the droplet initially moves along the rim of the Petri dish in a clockwise or counter-clockwise direction (Figure 5a,g). Importantly, the radius of these orbits is determined by the Petri dish, and not by the droplet. Remarkably, after a few hundred seconds, again in agreement with experiments, the droplet suddenly leaves the rim of the Petri dish and follows circular orbits with a characteristic radius and frequency (Figure 5b,h), which continue for several hundred seconds. Note that this transition from ballistic to chiral active motion does not require any external perturbations or changes in the system geometry, but happens automatically in the course of time in a perfectly reproducible way (after ~ 283 s on average, Figure S8). However, the detailed trajectories differ from each other in parallel experiments due to the randomness of the initial (and later) fluctuations. At even later times, yet another dynamical regime emerges, where chiral motion is limited to a small area of the Petri dish, in which the droplet is trapped (Figure 5c,i). Unlike the self-trapping phenomena reported in Refs. [60,61], where droplets are transiently localized inside a finite region of space, the self-trapping regime we report here refers to persistent chiral motion with essentially no net displacement during each period of oscillation.

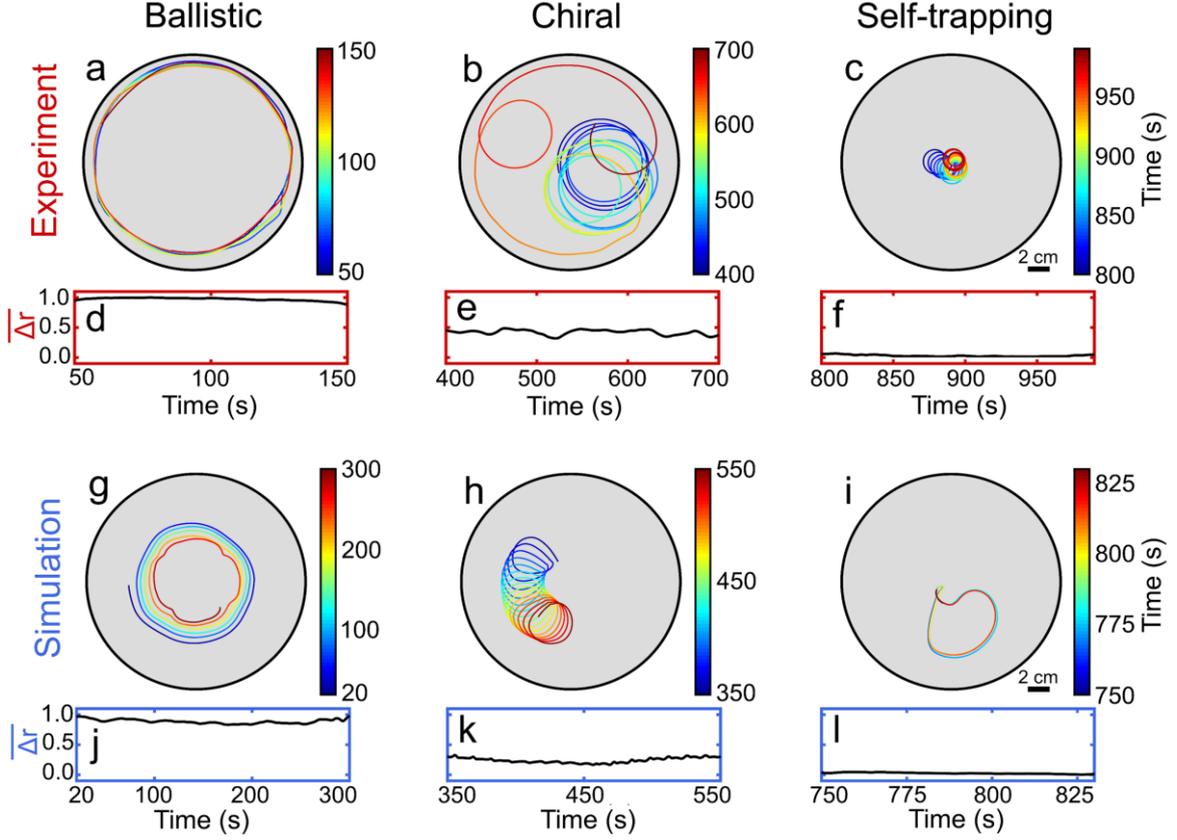

**Figure 5. Three dynamical regimes of droplet motion.** Typical droplet trajectories from experiments (a–c) and simulations (g–i). (d–f and j–l) Corresponding moving mean (averaged over 15 s) of the droplet displacement $\overline{\Delta r}$. We choose $\Delta t = 10$ s (d–f) and $\Delta t = 38$ s (j–l).

We now use our simulations, which offer direct access to the concentration profile and flow field at the air-water interface, to quantitatively characterize the different phases of motion observed both in experiments and simulations (Figure 5). To do so, we introduce the normalized displacement of the droplet within a fixed time interval $\Delta t$ as $\Delta r(t, \Delta t)=|\mathbf{r}(t+\Delta t)-\mathbf{r}(t)|/\max(|\mathbf{r}(t+\Delta t)- \mathbf{r}(t)|)$. The value of $\Delta t$ is chosen approximately as the average oscillation period of the droplet in the self-trapping regime, so that $\Delta r \sim 0$ in this regime. When the droplet is in the ballistic regime, it exhibits a maximum displacement within the time interval $\Delta t$, and hence $\Delta r \sim 1$ (Figure 5d,j). The chiral regime at intermediate time scales is characterized by a smaller value $\Delta r \sim 0.5$ (Figure 5e,k), whereas the late-time self-trapping regime corresponds to $\Delta r \sim 0$ (Figure 5f,l).

**2.5. Memory-induced chirality**

Our simulations show that the transition from ballistic to chiral active motion in self-solidifying droplet swimmers is caused by the interaction of the droplet with the PSS trail that it leaves behind while self-propelling. To see this, we now analyze the PSS distribution and flow field at different stages of the droplet motion. Initially, the droplet starts at the center of the Petri dish and moves towards the rim while leaving behind a small trail of PSS (Figure 6a). The trail is formed because PSS advection strongly dominates over diffusion, and so the emitted PSS follows the flow induced by the droplet motion. Following the high rate of PSS emission at early times, when the droplet encounters the rim of the Petri dish for the first time, PSS is advected and stays close to the rim. However, while moving along the rim, the droplet quickly speeds up and leaves behind a comparatively weak PSS trail. During this ballistic motion, the PSS concentration gradually builds up and hardly spreads diffusively, due to the small diffusion coefficient of PSS. The result is a continuously developing annular distribution of 'old' PSS with an additional peak around the position where the droplet encountered the rim for the first time (Figure 6b and Supplementary Movie S5).

Initially, the droplet dynamics are mainly determined by the interaction of the droplet with 'freshly' released PSS and the corresponding Marangoni flows at the droplet surface. However, at later times, as chemical production by the droplet weakens, the increasing concentration of 'old' PSS plays a growingly important role, especially at the edges (i.e., regions with a steep concentration gradient) of the annular region. As the droplet approaches these regions from the inner side, the overall flow (which includes contributions from both 'old' and 'fresh' PSS) makes the droplet turn back into the annular region. This leads to chiral motion within the annular concentration ring, the width of which determines the radius of the circular orbits (Figure 6b,c). (The concentration ring within which the droplet shows chiral motion fades over time, but a self-sustained domain of high chemical concentration remains, in which chiral motion persists.)

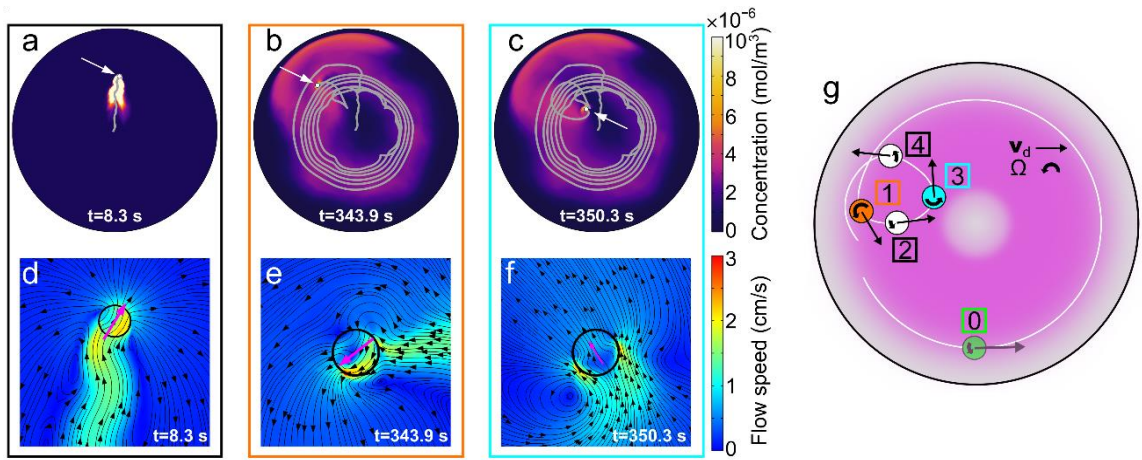

**Figure 6. Mechanism of chiral self-propulsion.** Simulation snapshots of the PSS concentration (a–c) and flow field (with streamlines) close to the droplet (d–f). (a–c) The white arrows point at the droplet position (white circle) and the gray lines show the droplet trajectory up to the time in the key. (d–f) The magenta arrows represent the droplet velocity. (g) Schematic of the droplet velocity $\mathbf{v}_d$ and average flow field vorticity $\Omega$ around the droplet (see *Methods* for the definition of $\Omega$) at representative instants of droplet motion indicated by numbers. The magenta region represents the ring of high PSS concentration where chiral motion sets in.

To better understand the transition from ballistic to chiral motion, we consider the droplet dynamics in the presence of a prescribed region of high chemical concentration (Figure 7a). If self-propulsion is fast enough, the droplet manages to completely traverse the accumulation of chemical (Figure 7b,d,f and Supplementary Movie S6). In contrast, for slow self-propulsion, the ambient flow prevails over self-propulsion, confining the droplet inside the high chemical concentration region and leading to chiral motion (Figure 7c,e,g and Supplementary Movie S7). In the latter case, the asymmetric flow field around the droplet makes the droplet turn whenever it tries to leave the circular domain (Figure 7e). In our experiments, it is the droplet itself that creates the annular region of high concentration giving rise to chiral motion (Figure 6a-c), which therefore constitutes a manifestation of memory effects in our system. Note that this transition to chiral motion for slow self-propulsion also takes place for a rectangular domain of high concentration (Figure 7h), further confirming that a specific confining geometry is not necessary to create chiral motion.

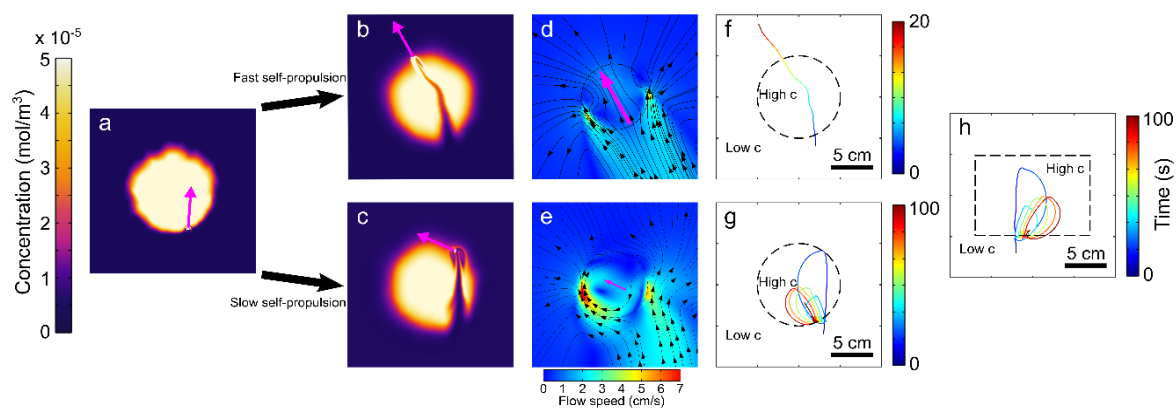

**Figure 7. Ballistic-to-chiral transition due to production rate decay.** (a-e) Simulation snapshots of the droplet (white circle) (a) approaching a circular region of high chemical concentration and moving through it for (b) fast and (c) slow self-propulsion, with the corresponding flow fields (with streamlines) close to the droplet (d,e). The magenta arrow indicates the instantaneous direction of droplet motion. (f,g) Corresponding droplet trajectories. The dashed lines represent the boundaries of the initial circular region, with a high chemical concentration inside and a low concentration outside. (h) Droplet trajectory in a rectangular region of high chemical concentration for slow self-propulsion.

How does the PSS trail change the direction of motion of the droplet? At early times, when the PSS concentration is much higher near the droplet boundary than in the bulk of the acidic water solution, the flow velocity decays fast with distance from the droplet boundary in all directions except behind the droplet (Figure 6d). In contrast, this decay is much slower at later times and the overall flow a few droplet radii away from the droplet boundary becomes comparatively strong (Figure 6e,f). Most importantly, the behavior of the average flow field vorticity $\Omega$, which is calculated as the curl of the flow field (see *Methods* for the definition of $\Omega$), changes notably in the surroundings of the droplet, as we quantitatively discuss in the next subsection. In particular, the vorticity increases significantly when the droplet is at the edges of the annular region. Accordingly, whenever the droplet tries to leave the annular domain, it experiences a strong tangential flow which turns it back into the annular domain (Figure 6e,f). Hence, the vorticity of the advective flow near the droplet, which is closely linked to the annular trail of 'old' PSS (Figure S14), is responsible for its transition from ballistic to persistent chiral motion, as schematically explained in Figure 6g. This mechanism of repetitive motion towards the edges of the annular domain and turning back into it creates a new route towards chiral motion. The key ingredient of this route is the ability of the droplet to respond to its own chemical history, rather than a specific symmetry in its environment, as illustrated by Figure 2e and S4.

**2.6. Correlation of memory-induced flow and droplet motion**

To understand the mechanism underlying chiral motion, we now consider $\theta_d$, the angle between the position vector of the droplet center and the *x*-axis (Figure 8a), which is a measure of the angular displacement of the swimmer. Initially, when the droplet moves ballistically around the center of the Petri dish, $\theta_d$ increases linearly with time until $\sim 300$ s (Figure 8b). After that, chiral dynamics come about characterized by the quasi-periodic oscillations between $\sim 350$ s and $\sim 650$ s, before entering the self-trapping phase.

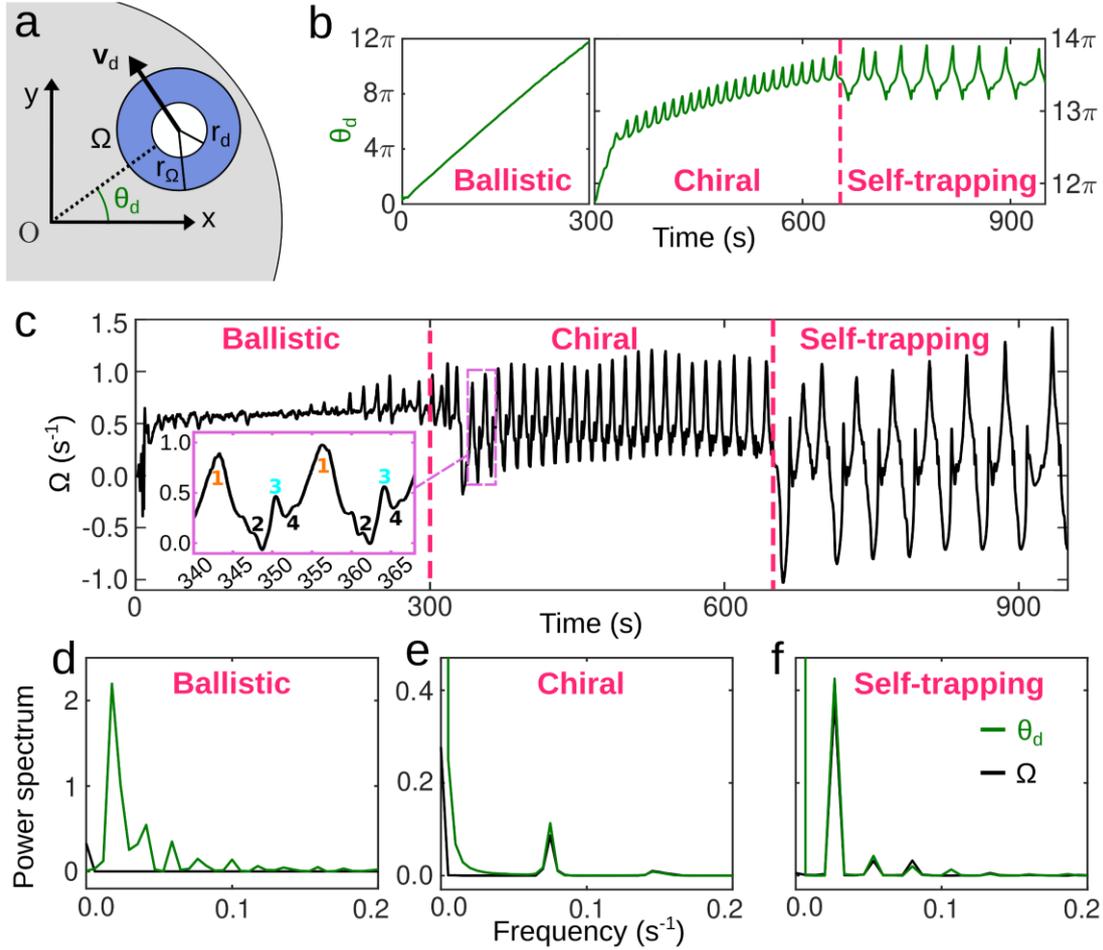

**Figure 8. Memory-induced advection leading to chiral dynamics.** (a) Schematic of the droplet swimmer (white circle) with radius $r_d$ and velocity $\mathbf{v}_d$ (black arrow) on the surface of water (grey background). $\Omega$ denotes the magnitude of the average flow field vorticity in the blue annular region of outer radius $r_\Omega = 4r_d$ (see *Methods* for the calculation of $\Omega$). The point O denotes the center of the Petri dish. (b,c) Time evolution of $\theta_d$ and $\Omega$ in the numerical simulation. The inset in (c) shows two oscillations in the chiral regime, with numbers corresponding to the instants shown in Figure 6g. (d–f) Power spectra (see *Methods*) of $\theta_d$ and $\Omega$, calculated using the data from the intervals 30–200 s, 400–600 s and 700–850 s respectively. To calculate the power spectra, we took into account the periodicity of $\theta_d$ by using values in the range $(-\pi, \pi]$.

To quantify the role of memory-induced advection, we now study the time evolution of $\Omega$, since a non-zero vorticity in the surrounding fluid breaks chiral symmetry (since the surrounding flow acquires a finite angular momentum). During the ballistic regime, $\Omega$ has a roughly constant value $\sim 0.5$ s$^{-1}$ (Figure 8c), which leads to circular droplet motion along the rim of the Petri dish at a constant angular velocity (Figure 8b). Once in the chiral regime, $\Omega$ is still non-zero, but it now exhibits quasi-periodic oscillations (Figure 8e). The local maxima of $\Omega$ in each oscillation (1 and 3 in Figure 8c, inset) occur when the droplet gets closest to the edges of the annular domain (see Figure 6g), with local minima (2 and 4) occurring in between. Overall, whenever the droplet reaches the edges of the annular landscape of the 'old' PSS, the strong vorticity of the local flow forces it to turn back into the annular region. A similar mechanism is observed in the self-trapping regime, where $\Omega$ still shows quasi-periodic oscillations (Figure 8f) but with its mean value close to 0 (Figure 8c), indicating that the droplet is confined, while still exhibiting (irregular) chiral dynamics.

The power spectra of $\theta_d$ and $\Omega$ (Figure 8d-f) clearly demonstrate the mechanism leading to the three distinct dynamical regimes. *While ballistic motion occurs when self-propulsion predominates over the effect of ambient flows, chiral motion and self-trapping occur when advection due to ambient flows prevails over self-propulsion* due to the low rate of PSS production. In fact, in each of the latter two regimes, both $\theta_d$ and $\Omega$ oscillate quasi-periodically with the same frequency, as shown by their almost identical power spectra (Figure 8e,f). Note that this crossover takes place gradually, rather than as a sharp transition, when the advection due to the ambient flows replaces Marangoni flows at the droplet surface as the driving force of motion. This gradual transition is memory-induced, since the ambient flows responsible for chiral motion are largely determined by the spatial distribution of the slowly diffusing PSS.

## 2.7. Self-propulsion in weak acidic water

Thanks to the rapid kinetics of polyelectrolyte complexation, the droplet swimmer is also viable in a weak acidic environment (pH 6.0). As shown in Figure 9a, we dipped a drop of the PEI/PSS solution into a reservoir filled with pH 1.25 water (left reservoir, diameter 4 cm), where a stable water-droplet interface forms within $\sim 180$ s due to PEI/PSS complexation. Then we adjusted the pH in the left reservoir to pH 5.1, before opening the gate between the two reservoirs to allow the droplet to move into the right reservoir (diameter 20 cm), of final pH 6.0. Figure 9b,c shows that, after swimming in the left reservoir for ~ 180 s, the droplet is able to maintain self-propulsion in the weak acidic water for ~ 10 min.

To control the motion of the droplet, superparamagnetic $Fe_3O_4$ nanoparticles were embedded into the droplet. As self-solidification proceeds, the droplet can be controlled by external magnetic fields to generate "HUST" and square-shaped trajectories (Figure 9d and 9e). Furthermore, an external PSS field generated by a PSS-releasing hydrogel can also be used to influence the trajectory of the droplet by inducing an exclusion zone in a certain time range (Figure S17).

We investigated the collective behavior of multiple droplets by dipping six droplets into a circular container (diameter 20 cm). As shown in Figure S18 and Movie S8, the droplets initially perform random ballistic motion and repel each other upon contact due to the (repulsive) diverging flow. Later on, as the PSS release rate decreases, memory effects and capillary attraction start to play a role, and the droplets start to approach each other (Figure S18c), eventually forming a cluster (Figure S18d).

## 2.8. Uranium removal

As a proof-of-concept demonstration, the droplet swimmer can act as an advanced "cleaning robot" for the removal of radioactive uranium from water, based on sulfonate-uranyl coordination (Figure 9f). Droplets were first complexed in pH 1.25 water for 3 min, 10 min and 6 h, respectively, and then moved to a uranium solution (10 ppm). The 1$^{st}$ droplet (complexed for 3 min) shows the fastest removal kinetics, with 84.7% of the uranium removed within 90 min, whereas the 2$^{nd}$ (complexed for 10 min) and 3$^{rd}$ (complexed for 6 h) droplets remove 54.1% and 8.7% of the uranium within the same time (Figure 9g). The 1$^{st}$ droplet has a higher velocity (Figure S19a) and explores a larger area than the 2$^{nd}$ droplet (Figure 9h) within the first 20 min of motion, while the 3$^{rd}$ droplet is static. As one would expect, self-propulsion allows the droplet to explore regions where uranyl ions have not been adsorbed. Moreover, the flow created by the droplet facilitates mixing of the solution by breaking the passive diffusion of uranyl ions. Therefore, the efficiency of uranium removal is greatly enhanced by the active motion of the droplet swimmer. Furthermore, the uranium removal efficiency of the 1st droplet increases as the molar ratio of PSS to PEI increases (Figure S20). The full scan X-ray photoelectron spectroscopy (XPS) spectra before and after adsorption indicate that uranium indeed binds to the droplet (Figure 9i). The high-resolution U4f spectrum confirms that the valence state of uranium does not change after adsorption (Figure 9j). Furthermore, the FTIR spectrum of the droplet shows an additional peak at 892.3 cm$^{-1}$ after adsorption (Figure S19b), corresponding to the vibration of O=U=O. Note that when the concentration or volume of uranium-wasted water is high, the removal efficiency of the limited number of droplets decreases (Figure S19c and d). Therefore, a large number of droplets is needed to reach a high removal efficiency. Moreover, the presence of a large number of organic pollutants/solvents, which influences the interface stability of the droplet and the distribution of PSS, could affect the adsorption performance of the droplet.

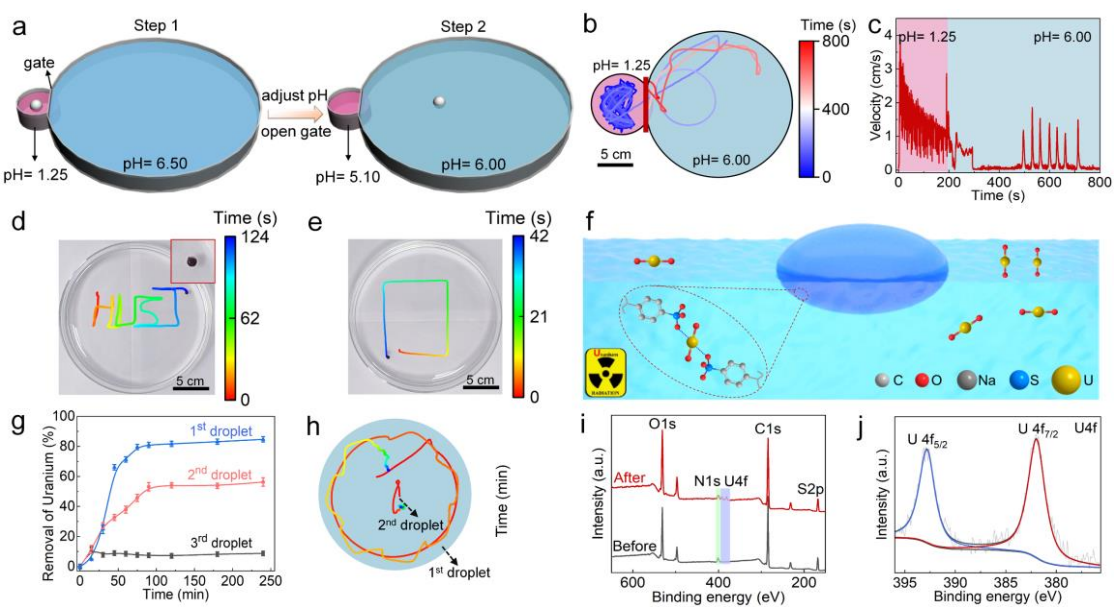

**Figure 9. Applications of the droplet swimmer.** (a) Schematic of the setup used to achieve droplet self-propulsion in weak acidic water (right cell) after formation of a stable interface in strong acidic water (left cell). See Figure S15 for the dimensions of the setup. (b) Trajectory and (c) velocity of the droplet swimmer in pH 1.25 and 6.00 water. Trajectories of a droplet with embedded Fe$_3$O$_4$ nanoparticles controlled by a magnet: (d) "HUST" and

(e) square shapes. (f) Schematic of the removal of uranium from wastewater by the droplet swimmer. (g) Uranium removal kinetics for droplets complexed in pH 1.25 water for different times (1st droplet: 3 min, 2nd droplet: 10 min, 3rd droplet: 6 h). (h) Trajectory of the 1st and 2nd droplets in a uranium solution of 10 ppm. (i) Full scan XPS, (j) high-resolution U4f XPS of the 1st droplet swimmer before and after adsorption of uranium.

## 3. Conclusion

We have introduced, realized and analyzed the first self-solidifying droplet swimmer, which makes use of a slowly proceeding solidification transition to power self-propulsion. Unlike most existing synthetic swimmers, self-solidifying droplets do not require any persistent external fueling; rather, they are intrinsically equipped with an internal energy depot. As one of their key features, these swimmers show a *dynamical transition* to chiral motion without requiring any explicit symmetry breaking or a complex (viscoelastic) environment. Our results establish an alternative mechanism for fueling self-propulsion which can be deployed in other systems, such as hydrogels. They also unveil a so far unknown relation between chirality and memory effects which works even in simple environments that shape the swimmer trails such that the swimmer becomes trapped by its own chemical history. Moreover, the droplet swimmer is capable of a highly efficient uranium removal from wastewater. This self-powering approach could be used to optimize next-generation microswimmers for tasks like optimal surface coverage, targeted drug delivery or microsurgery in environments where external refueling is not easily possible. Potential methods of miniaturizing the droplet to the microscale include microfluidics and membrane emulsification techniques.[62,63]

## 4. Methods Section

**Experimental setup**

*Materials:* Poly(sodium 4-styrenesulfonate) (PSS, average molecular weight ~ 1,000,000 and ~ 70,000) was purchased from Sigma-Aldrich (USA). Polyethylenimine (PEI, branched, linear formula: $(CH_2CH_2NH)_n$, average molecular weight ~ 70,000, 50 wt% in $H_2O$), $Fe_3O_4$ nanoparticle suspension and $UO_2(NO_3)_2 \cdot 6H_2O$ were purchased from Shanghai Aladdin Biochemical Technology Co., Ltd (China). Ethanol (anhydrous, purity of 99.5 %), HCl, $Na_2CO_3$, concentrated nitric acid and NaCl (purity of 99.5 %) were obtained from Sinopharm Chemical Reagent Co., Ltd (China). Deionized water was purified by a ULUP water purification system with a minimum resistivity of 18.25 m$\Omega$ cm$^{-1}$ and used in all the processes. A stock solution of U(IV) (1000 mg L$^{-1}$) was prepared by dissolving $UO_2(NO_3)_2 \cdot 6H_2O$ in concentrated nitric acid (2 mL) and deionized water (98 mL). The stock solution was then diluted to a desired concentration prior to the uranium adsorption experiment with the pH being adjusted by $Na_2CO_3$.

*Preparation of the PEI/PSS solution*: We prepared a PSS and PEI mixture solution of 25 wt% solids content. Typically, 2.06 g PSS powder and 0.86 g PEI solution (50 wt%) were added into 9.53 mL of deionized water and mixed thoroughly on a magnetic stirrer for 6 h to obtain a maize-yellow solution. To study the influence of the molecular weight of PSS on the motion of the droplet, a PEI/PSS solution of 25 wt% solids content and PSS of molecular weight ~ 70,000 was prepared in the same way. To investigate the influence of solids content and the PSS-to-PEI molar ratio on the self-propulsion behavior of the droplet, solutions of various solids contents and PSS-to-PEI ratios were prepared in the same way. For magnetic control, $Fe_3O_4$ nanoparticle suspension was mixed with PEI/PSS solution of 25 wt% solids content for droplet preparation.

*Motion of the droplet swimmer*: A droplet of the PEI/PSS solution of the desired volume was deposited onto an acidic solution of pH 1.25 in a Petri dish (diameter of 20 cm and height of 15 mm) from a fixed height of ∼ 1 cm above the air-water interface. The droplet starts self-propulsion at the air-water interface as it slowly solidifies from the outer edge to the inside (see Figure 1d). The motion of the droplet swimmer was recorded by a digital camera at a frame rate of 30 Hz and its position in each frame was obtained by extracting the perimeter using a self-written Python script,[64] from which the trajectory and velocity of the swimmer were calculated.

To visualize the fluid flow generated by the droplet pump, a needle was used to fix the droplet at the air-water interface and polystyrene particles of diameter 36 μm (Microparticles, GmbH) were used as tracer particles. The motion of the tracer particles was recorded by a digital camera at a frame rate of 30 Hz from the top and side views, respectively. Videos were analyzed by tracking the position of the tracer particles in each frame via Tracker.

*Uranium removal experiment:* The uranium removal experiment was performed at room temperature. Fifteen droplet swimmers were dipped into strong acidic water (pH 1.25) for different times (3 min, 10 min and 6 h). Then they were moved to uranium solutions of 10 ppm (pH 4.3, 10 mL). Specimens were taken at different times to determine the remaining uranium concentration from their UV-Vis absorption spectra (see details in Supplementary Note 9). The adsorption percentage of uranium was calculated as: $(C_0-C_t)/C_0\times 100$, where $C_0$ and $C_t$ are the initial concentration of uranium and the concentration at time $t$.

*Characterization:* The morphology and microstructure of the droplet were observed by field-emission scanning electron microscopy (Quanta 200). Samples were prepared by fast freezing the droplet in liquid nitrogen followed by freeze-drying in a freeze drier to keep the morphology. The concentration of PSS released by the droplet was monitored by taking small specimens of acidic water (∼ 0.5 mL) from the surface of the solution at different positions with respect to the center of the droplet and detected on a UV-Vis spectrophotometer (SolidSpec-3700, Shimadzu) at 255 nm. The evolution of the surface tension gradients generated by the droplet was measured in-situ with a force tensiometer using a thin plate geometry of 10 mm in width and 5 mm in height (Sigma 700, Biolin Scientific). The surface tension of the acidic water solution was measured by the pendant drop method on a micro-optical contact angle meter (OCA15EC, Dataphysics). The Fourier-transform infrared spectroscopy (FTIR) spectra of PEI, the PEI/PSS mixture, the solidified droplet and the droplet after adsorption of uranium were characterized on INVENIO-R, Bruker, Germany. Dynamic light scattering (Nano ZS 90, Malvern) was used to characterize the diffusion coefficients of PSS and PEI in the dilute concentration range (about 0.5 mg

mL$^{-1}$). The pH of the acidic water solution was measured by a pH meter at room temperature (PB-10, Sartorius). The surface element composition of the droplet before and after uranium adsorption was characterized by XPS (AXIS-ULTRA DLD-600W).

**Model.** We model the droplet swimmer as a two-dimensional circular disk of fluid of radius $r_d$ floating on the surface of another fluid (i.e., the acidic water solution), which is confined within a bigger circular domain (i.e., the Petri dish) of radius $R$ corresponding to the radius of the Petri dish used in the experiments. The droplet emits polymer (PSS) molecules from its boundary into the surrounding fluid at an initial production rate $A$, and the PSS concentration $c(r, t)$ evolves according to the advection-diffusion equation

$$\frac{\partial c}{\partial t} = D\nabla^2 c - \mathbf{u}_w \cdot \nabla c, \quad (1)$$

where $D$ denotes the diffusion coefficient of the polymer and $\mathbf{u}_w(\mathbf{r}, t)$ is the flow velocity of the acidic water solution. PSS production is modeled by an isotropic time-dependent flux $\mathbf{n} \cdot (D\nabla c)|_S = Ae^{-t/\tau}$ (see Figure 2b and S10) at the droplet boundary $S$. $\mathbf{n}$ denotes the outward-pointing normal vector to the droplet boundary and $\tau$ is the characteristic time scale of PSS emission. By analogy with the experiments, the surface tension $\sigma(\mathbf{r}, t)$ of the acidic water solution decreases due to the emitted polymer molecules (Figure S7), which we model via the relation

$$\sigma(\mathbf{r},t) = \frac{\sigma_0}{1+\dfrac{k}{\sigma_0}c(\mathbf{r},t)}, \quad (2)$$

where $\sigma_0$ is the surface tension of water in the absence of polymer molecules and $k$ is the surface tension coefficient. Fluid flows are modeled by the incompressible unsteady Stokes equations

$$\begin{aligned}
\rho_w \frac{\partial \mathbf{u}_w}{\partial t} &= \mu_w \nabla^2 \mathbf{u}_w - \nabla p_w + \mathbf{F}_w, \\
\rho_d \frac{\partial \mathbf{u}_d}{\partial t} &= \mu_d \nabla^2 \mathbf{u}_d - \nabla p_d, \quad (3)\\
\nabla \cdot \mathbf{u}_{w,d} &= 0.
\end{aligned}$$

Here, the subscripts $w$ and $d$ denote the acidic water solution and the droplet respectively. $\rho_{w,d}$, $\mathbf{u}_{w,d}$, $\mu_{w,d}$ and $p_{w,d}$ denote, respectively, the density, flow velocity, dynamic viscosity and pressure of the corresponding medium. The hydrodynamic drag forces (advection) resulting from Marangoni flows caused by a spatially asymmetric polymer distribution at the air-water interface are taken into account by introducing the *bulk Marangoni force* $\mathbf{F}_w = k_b \nabla \sigma$, where $k_b$ is the *bulk surface tension coefficient*. Additionally, a surface tension gradient along the droplet boundary creates a *direct Marangoni force*, which is accounted for by the stress boundary condition for the Stokes Equations 3 at the droplet boundary:[65]

$$\mathbf{n} \cdot (\mathbf{T}_w - \mathbf{T}_d)|_S = \sigma(\nabla_S \cdot \mathbf{n})\mathbf{n}|_S - \nabla_S \sigma|_S . \quad (4)$$

where $\mathbf{T}_{w,d} = -p_{w,d}\mathbf{I} + \mu_{w,d}(\nabla \mathbf{u}_{w,d} + (\nabla \mathbf{u}_{w,d})^T)$ is the stress tensor, with $\mathbf{I}$ the identity matrix, and $\nabla_S$ denotes the gradient operator along the droplet boundary $S$. Additionally, a no-slip boundary condition is imposed at the circular boundary of the acidic water domain. We define the velocity $\mathbf{v}_d(t)$ of the droplet swimmer as the average flow velocity of the fluid inside the droplet domain, i.e., $\mathbf{v}_d(t) = \langle \mathbf{u}_d(\mathbf{r}, t)\rangle_{\text{droplet}}$.

Figure S7 explicitly shows that the flow profile created by the droplet is three-dimensional. However, rather than a 3D model, here we use a 2D model which effectively describes the processes taking place at the air-water interface, while remaining numerically feasible. This model successfully grasps the linear decay of the flow with increasing radial distance to the droplet (Figure 3c). An accurate description of the flow profile close to the droplet, which mainly has vertical contributions (see Figure S7), is nonetheless beyond the scope of our model.

**Definitions.** *Vorticity*: The average vorticity $\Omega$ is calculated over an annular region of inner radius $r_d$ and outer radius $r_\Omega = 4r_d$ centered about the droplet position $\mathbf{r}_{\text{droplet}}$, i.e., $\Omega \hat{z} = \langle \nabla \times \mathbf{u}_w(\mathbf{r}, t)\rangle_{r_d < |\mathbf{r} - \mathbf{r}_{\text{droplet}}| < r_\Omega}$, where $\hat{z}$ is the upward-pointing unit vector perpendicular to the air-water interface.

*Power spectra*: We define the power spectra $S(\omega_k)$ of $\theta_d(t_n)$ and $\Omega(t_n)$ as $S(\omega_k) = 1/N^2 |F(\omega_k)|^2$, where $N$ and $F(\omega_k)$ are, respectively, the number of data points and the fast Fourier transform of the corresponding quantity, i.e.,

$$F(\omega_k) = \sum_{n=0}^{N-1} \theta_d(t_n) e^{-i2\pi kn/N} \text{ or } F(\omega_k) = \sum_{n=0}^{N-1} \Omega(t_n) e^{-i2\pi kn/N}. \quad (5)$$

**Simulations**

*Software and numerical scheme*: We used the COMSOL Multiphysics® software, which is based on the finite element method (FEM), to model our setup and numerically solve Equations 1-4. Only Equations 3 and 4 were solved inside the droplet domain, whereas all the equations were solved in the acidic water domain. We used a direct solver, with the Backward Differentiation Formula (BDF) time-stepping algorithm. A triangular mesh with linear elements was used, with element size increasing from the boundary of the droplet towards the bulk of the acidic water domain, thus ensuring numerical accuracy at positions where the chemical concentration is expected to be higher due to emission. Additionally, we included 30 boundary layers at the outer boundary of the acidic water domain, i.e., at the rim of the Petri dish.

*Simulation parameters:* We chose the simulation parameters based on the experimental conditions: $r_d = 0.22$ cm, $D = 8$ µm$^2$ s$^{-1}$ (see Supplementary Note 3), $\sigma_0 = 72$ mN m$^{-1}$, $\rho_w = \rho_d = 10^3$ kg m$^{-3}$, $\mu_w = 10^{-3}$ Pa s. In the experiments, the viscosity inside the droplet increases over time due to solidification. However, since the inner viscosity is three orders of magnitude higher than that of water (Figure S9), using a time-dependent $\mu_d$ would not affect our simulations significantly. Accordingly, for simplicity, we use a constant value $\mu_d = 1$ Pa s. $R = 10$ cm, except in Figure 3d and 7, for which both the experiment (3D) and the simulations (2D) were conducted in a tank of acidic water of sides $100 \times 60$ cm to mimic bulk conditions. $A = 2/3 \times 10^{-7}$ mol m$^{-2}$ s$^{-1}$ and $\tau^{-1} = 0.11$ min$^{-1}$ were obtained from fits to the experimental data (see Figure 2b and Supplementary Note 3). To produce Figure 3d and S13, we added a weak excess activity $A_e = 10^{-8}$ mol m$^{-2}$s$^{-1}$ on one half of the droplet boundary for 0.01 s to reinforce the spontaneous symmetry breaking triggering self-propulsion. Note that changes in the value of $A_e$ hardly affect our results, and therefore our simulations are robust to the addition of a weak initial asymmetry. In Figure 7, Supplementary Movie S6 and S7, we set an initial droplet velocity $\mathbf{v}_d = 2\hat{y}$ cm s$^{-1}$, and an initial production rate $0.53 \times A$ (Figure 7b,d,f and Supplementary Movie S6), which corresponds to the production rate at $t = 350$ s in the experiments (Figure 2b and S10), and $0.053 \times A$ (Figure 7c,e,g,h and Supplementary Movie S7). To match the maximum droplet speed in bulk conditions to the experimental one, we set $k = 1$ N m$^2$ mol$^{-1}$ and $k_b = 10^5$ m$^{-1}$ in Figure 3c,d and S13. The position of the maximum and the shape of the curve then follow from this choice of parameter values. We then set $k = 10$ N m$^2$ mol$^{-1}$ and $k_b = 400$ m$^{-1}$ in all other Figures, Supplementary Figures and in Supplementary Movie S5-S7. These parameters yield a value of the Péclet number of Pe = $v_d r_d/D \sim 10^7$.

## Acknowledgments

This work is financially supported by the National Natural Science Foundation of China (No. 22832059 and 22178139), Huazhong University of Science and Technology (No.2021XXJS036). We are grateful to the Analytical and Testing Centre of HUST for access to their facilities. A. K. M. and B. L. acknowledge financial support by the Deutsche Forschungsgemeinschaft (DFG, German Research Foundation) through project number 233630050 (TRR-146).

## Author contributions

[#]These authors contributed equally to this work. R.N., Q.Z., and B.L. initiated and designed the project. R.N. and K.F. performed the experiments and analyzed the data; J.C.U.M. and A.K.M. conducted the simulations and analyzed the data. R.N., J.C.U.M., A.K.M. and B.L. participated in the writing of the manuscript, with input from K.F and J.P.Q. All authors approved the final version of the manuscript.